\documentclass[structabstract]{aa} 
\usepackage{float}
\usepackage{soul}   
\bibpunct{(}{)}{;}{a}{}{,} 
\usepackage{natbib}
\usepackage{epstopdf}
\usepackage{graphicx}

\begin{document}

   \title{Testing the two planes of satellites in the Centaurus group}

   \author{Oliver M\"uller
          \inst{1}
          \and
          Helmut Jerjen
          \inst{2}
          \and
          Marcel S. Pawlowski\inst{3}
          \and
          Bruno Binggeli \inst{1}
          }

   \institute{Departement Physik, Universit\"at Basel, Klingelbergstr. 82, CH-4056 Basel, Switzerland\\
         \email{oliver89.mueller@unibas.ch; bruno.binggeli@unibas.ch}
                  \and
         Research School of Astronomy and Astrophysics, Australian National University, Canberra, ACT 2611, Australia\\
         \email{helmut.jerjen@anu.edu.au}
                           \and
         Department of Astronomy, Case Western Reserve University, 10900 Euclid Avenue, Cleveland, OH 44106, USA \\
         \email{marcel.pawlowski@case.edu}
             }

   \date{Received 13 July 2016; accepted 24 August 2016}

 
  \abstract
   {The existence of satellite galaxy planes poses a major challenge for the standard picture of structure formation with non-baryonic dark matter. Recently Tully et al.~(2015) reported the discovery of two almost parallel planes in the nearby Cen\,A group using mostly high-mass galaxies (M$_B$\,<\,-10\,mag) in their analysis.}
   {Our team detected a large number of new group member candidates in the Cen\,A group. This dwarf galaxy sample, combined with other recent results from the literature, enables us to test the galaxy distribution in the direction of the Cen\,A group and to determine the statistical significance of the geometric alignment.}
   {Taking advantage of the fact that the two galaxy planes lie almost edge-on along the line of sight, the newly found group members can be assigned relative to the two planes. We used various statistical methods to test whether the distribution of galaxies follows a single normal distribution or shows evidence of bimodality as has been reported earlier.}
   {We confirm that the data used for the Tully et al.~(2015) study support the picture of a bimodal structure. When the new galaxy samples are included, however, the gap between the two galaxy planes is closing and the significance level of the bimodality is reduced. Instead, the plane that contains Cen\,A becomes more prominent. 
   }
{We found evidence that the galaxy system around Cen\,A is made up of only one plane of satellites. This plane is almost orthogonal to the dust plane of Cen\,A. Accurate distances to the new dwarf galaxies will be required to measure the precise 3D distribution of the galaxies around Cen\,A.}


   \maketitle
%

\section{Introduction}

The mere abundance and spatial distribution of faint dwarf galaxies provide a powerful testbed for dark matter and structure formation models on Mpc and galaxy scales. The standard picture of structure formation with dark matter is heavily challenged by the highly asymmetric features found in the distributions of dwarf galaxies in the Local Group, {which was first noted by \cite{2005A&A...431..517K}}. There is the vast polar structure \citep[VPOS;][]{2012MNRAS.423.1109P,2015MNRAS.453.1047P,2016MNRAS.456..448P}, a thin (rms 
height $\approx$ 30 kpc) highly inclined, co-rotating substructure of faint satellite galaxies, young globular clusters, and stellar streams, spreading in Galactocentric distance between 10 and 250\,kpc. A similar feature was found in the Andromeda galaxy surroundings, the so-called Great Plane of Andromeda \citep[GPoA;][]{2006AJ....131.1405K,2007MNRAS.374.1125M,2013Natur.493...62I}. On a {slightly} larger scale, two dwarf galaxy planes containing all but one of the 15 non-satellite galaxies have been identified in the Local Group \citep{2013MNRAS.435.1928P}. Such extreme satellite planes are found in only  $<0.1\%$ of simulated systems in cosmological simulations \citep[e.g.,][]{2014ApJ...784L...6I,2014MNRAS.442.2362P}, making them difficult to accommodate in a standard $\Lambda$CDM scenario. An alternative analysis of cosmological simulations, based on including the look elsewhere effect but ignoring observational uncertainties and the non-satellite planes \citep{2015MNRAS.452.3838C}, finds that only about 1 per cent of Local Group-equivalent environments should host similarly extreme satellite structures.

The fundamental question arises whether the relative sparseness and asymmetric distribution of low-mass dwarf galaxies encountered in the Local Group is a statistical outlier or a common phenomenon in the local universe. \citet{2014Natur.511..563I} have approached this question with a statistical study of velocity anticorrelations among pairs of satellite galaxies on opposite sides of their host, using data from the SDSS survey. They found a strong excess in anticorrelated velocities, which is consistent with co-orbiting planes, but their conclusions were based on a small sample of only 22 systems, and have been challenged since \citep[][{ but see \citealt{2015ApJ...805...67I}}]{2015MNRAS.453.3839P,2015MNRAS.449.2576C}. A different approach to address the question of satellite planes is to extend searches for such structures to satellite populations in other galaxy groups in the nearby universe. For example, \citet{2013AJ....146..126C} have found that dwarf spheroidal galaxies in the M81 group lie in a flattened distribution.

Most recently, \cite{2015ApJ...802L..25T}, hereafter T15, reported evidence for a double-planar structure around Centaurus\,A (Cen\,A, NGC5128) in the nearby Centaurus group of galaxies, with properties reminicent of the two Local Group dwarf galaxy planes \citep{2013MNRAS.435.1928P}. 
Furthermore, \cite{2015MNRAS.452.1052L} found that the Local Group and the Cen\,A group reside in a filament stretched by the Virgo Cluster and compressed by the Local Void and smaller voids. Four out of five planes of satellites (including the two galaxy planes around Cen\,A) align with this filament with the normal vectors pointing in {the} direction of the Local Void and the planes almost parallel to the minor axis of the filament.
These results demonstrate that systematic studies of the spatial distribution of low luminosity galaxies in nearby groups can provide important observational constraints for further testing of structure formation models outside of the Local Group.

The aim of the present study is to test how recently discovered dwarf galaxies in the Centaurus group \citep{2014ApJ...793L...7S, 2014ApJ...795L..35C,2016ApJ...823...19C, 2015A&A...583A..79M,2016arXiv160504130M} are distributed in the double planar structure reported by Tully and collaborators. Comparing the footprint of the two planes, 20 of these dwarf candidates are located in that region of the sky. It is important to note that this analysis is feasible without distance information because of special geometry: the normal vectors of the two planes are almost parallel and perpendicular to the line of sight. Consequently, any distance uncertainties for the new galaxies does not move them in or out of the two planes.

In section 2 we present the four different galaxy samples used in our analysis. In section 3 we show the transformation between the equatorial coordinate system to the Cen\,A reference frame and fit the two planes of satellites. Section 4 follows a discussion of the geometrical alignment of the planes and the distribution of galaxies. We test the statistical significance of the two planes in Section 5, followed by Section 6 where we summarize the results.  

\section{Sample description}
In recent years several untargeted imaging surveys were dedicated to search for low surface brightness dwarf galaxies in the nearby universe \citep[e.g.,][]{2009AJ....137.3009C,2013AJ....146..126C,2014ApJ...787L..37M,2016A&A...588A..89J}.  In particular, the richest galaxy aggregate in the Local Volume, the Centaurus\,A (Cen\,A) group of galaxies received attention. The Panoramic Imaging Survey of Centaurus and Sculptor \citep[PISCeS;][]{2014ApJ...793L...7S, 2014ApJ...795L..35C,2016ApJ...823...19C} revealed 13 extremely faint dwarf galaxies in the vicinity ($\sim11$\,deg$^2$) of Cen\,A.  Group memberships of nine dwarfs have been confirmed with the tip of the red giant branch (TRGB) method. Our team conducted a survey of 550 square degrees around the Centaurus group, including the Cen\,A and M\,83 subgroups, discovering 57 potential group member dwarf galaxies \citep{2015A&A...583A..79M,2016arXiv160504130M}. 

The Centaurus group is the largest concentration of galaxies in the Local Volume (Distance < 10\,Mpc). Before our study there were about 60 group members known \citep{2004AJ....127.2031K,2013AJ....145..101K}. Similar to the Local Group, the Centaurus group has two gravitational centers consisting of a larger galaxy population around the massive, peculiar galaxy NGC\,5128 (Cen\,A) at a mean distance of 3.8\,Mpc, and a smaller concentration around the giant spiral M\,83 at a mean distance of 4.9\,Mpc \citep{2004AJ....127.2031K,2013AJ....145..101K, 2015ApJ...802L..25T, 2015AJ....149..171T}.

This work makes use of four different galaxy samples; these are listed in Table\,\ref{table:1}. The first sample (1), hereafter called {T15 sample}, is almost identical to the sample of T15 as given in their Table\,1. The authors subdivided their galaxies into six subsamples: Plane 1, Plane 2, Plane 1?, Plane 2?, other and other?. The Plane 1 subsample consists of 14 galaxies. Following T15, we exclude the extremely faint dwarf galaxies Dw-MM-Dw1 and Dw-MM-Dw2 \citep{2014ApJ...795L..35C} to avoid any selection bias. The Plane 2 subsample contains 11 galaxies. In the other four subsamples six galaxies lack measured distance information, meaning that they are considered Cen\,A members based on morphological and/or surface brightness grounds. We exclude PGC\,45628 from these because its projected distance from Cen\,A is larger than 1\,Mpc.
Three other galaxies (ESO\,219-010, ESO\,321-014, and PGC\,51659) have measured distances but cannot be unambiguously assigned to one of the two planes; we exclude these galaxies from the sample. In summary, our {T15 sample} contains 25 galaxies that have measured distances (members) and five without (candidates).

The second sample (2) includes the 25 members from the {T15 sample} and the nine newly discovered dwarf galaxy members from \cite{2014ApJ...795L..35C,2016ApJ...823...19C}, including the two PISCeS dwarfs that were excluded in the first sample. Hereafter we call it the {LV sample} because all of these 34 galaxies with distances are part of the Local Volume (LV) sample listed in the online version of the Updated Nearby Galaxy Catalog \citep{2013AJ....145..101K}.

The third sample (3) comprises all candidate members of the Cen\,A subgroup known to date between $197.5^{\circ}<\alpha_{2000}<207.5^{\circ}$ and $-46^{\circ}<\delta_{2000}<-36^{\circ}$. This sample, called hereafter the {Candidate sample}, consists of 25 dwarf candidates in the vicinity of Cen\,A, including the new dwarf candidates from \cite{2016arXiv160504130M} and the four candidates without distances from \cite{2016ApJ...823...19C}. All of these candidates are considered likely members of the Cen\,A subgroup based on morphological and/or surface brightness grounds.

The fourth sample (4) contains all Cen\,A group members and candidates known to date, hereafter called the {Complete sample}. It is the combination of the 34 galaxies with distances ({LV sample}) and the 25 candidates from the {Candidate sample}. The {LV sample} and the {Complete sample} contain only galaxies with radial distances smaller than 1\,Mpc from Cen\,A (3.68\,Mpc). 
See Table\,\ref{table:2} for a complete list of galaxies (names, coordinates, and distances) included in our analysis.

\begin{table}
\caption{Galaxy numbers in the four samples. Members and candidates are galaxies with and without distance measurements, respectively. }             
\label{table:1}      
\centering                          
\begin{tabular}{l c c c}        
\hline\hline            
 & Members & Candidates & Total\\    
Sample name & (N) & (N) & (N)\\    
\hline     \\[-2mm]                   
   T15 sample (1)& 14+11& 5 &30 \\      
   LV sample (2) & 34 & 0    & 34 \\
   Candidate sample (3) & 0 & 25    & 25 \\
   Complete sample (4) & 34 & 25    & 59 \\
\hline                                   
\end{tabular}
\end{table}

\begin{table*}
\caption{Members and possible members of the Cen\,A subgroup.}
\label{table:2}      
\begin{minipage}{.5\linewidth}
\centering                          
\begin{tabular}{l c c l l}        
\hline\hline                 
 & $\alpha_{2000}$ & $\delta_{2000}$ & $D$ & \\    
Galaxy Name & (deg) & (deg) &  (Mpc) & Sample\\    
\hline      \\[-2mm]                  
ESO269-037$^1$        &         195.8875&       -46.5842&   3.15&(1)(2)(4)\\
NGC4945$^1$           &         196.3583&       -49.4711&       3.72&(1)(2)(4)\\
ESO269-058$^1$        &         197.6333&       -46.9908&       3.75&(1)(2)(4)\\
KKs53$^1$             &         197.8083&       -38.9061&       2.93&(1)(2)(4)\\
KK189$^1$             &         198.1875&       -41.8319&       4.23&(1)(2)(4)\\
ESO269-066$^1$        &         198.2875&       -44.8900&       3.75&(1)(2)(4)\\
NGC5011C$^1$          &         198.2958&       -43.2656&       3.73&(1)(2)(4)\\
KK196$^1$             &         200.4458&       -45.0633&       3.96&(1)(2)(4)\\
NGC5102$^1$           &         200.4875&       -36.6297&       3.74&(1)(2)(4)\\
KK197$^1$             &         200.5042&       -42.5356&       3.84&(1)(2)(4)\\
KKs 55$^1$                &             200.5500&       -42.7308&       3.85&(1)(2)(4)\\
NGC5128$^1$           &         201.3667&       -43.0167&       3.68&(1)(2)(4)\\
KK203$^1$             &         201.8667&       -45.3525&       3.78&(1)(2)(4)\\
ESO324-024$^1$        &         201.9042&       -41.4806&       3.78&(1)(2)(4)\\
NGC5206$^2$           &         203.4292&       -48.1511&       3.21&(1)(2)(4)\\
NGC5237$^2$           &         204.4083&       -42.8475&       3.33&(1)(2)(4)\\
NGC5253$^2$           &         204.9792&       -31.6400&       3.55&(1)(2)(4)\\
KKs 57$^2$            &         205.4083&       -42.5819&       3.83&(1)(2)(4)\\
KK211$^2$             &         205.5208&       -45.2050&       3.68&(1)(2)(4)\\
KK213$^2$             &         205.8958&       -43.7691&   3.77&(1)(2)(4)\\
ESO325-011$^2$        &         206.2500&       -41.8589&       3.40&(1)(2)(4)\\
KK217$^2$             &         206.5708&       -45.6847&       3.50&(1)(2)(4)\\
CenN$^2$              &         207.0375&       -47.5650&       3.66&(1)(2)(4)\\
KK221$^2$             &         207.1917&       -46.9974&       3.82&(1)(2)(4)\\
ESO383-087$^2$        &         207.3250&       -36.0614&       3.19&(1)(2)(4)\\
KK198                 &         200.7342&   -33.5728&   3.68*&(1)(3)(4)\\
KKs54                 &         200.3850&       -31.8864&       3.68*&(1)(3)(4)\\
KKs59                 &         206.9920&       -53.3476&       3.68*&(1)(3)(4)\\
KKs58                 &         206.5042&       -36.3281&       3.68*&(1)(3)(4)\\
KKs51                 &         191.0896&       -42.9397&       3.68*&(1)(3)(4)\\
\end{tabular}
 \end{minipage}%
    \begin{minipage}{.5\linewidth}
\begin{tabular}{l c c l l}        
\hline\hline                 
 & $\alpha_{2000}$ & $\delta_{2000}$ & $D$ & \\    
Galaxy Name & (deg) & (deg) &  (Mpc) & Sample\\    
\hline      \\[-2mm]                        
CenA-MM-Dw5       &             199.9667&       -41.9936&       3.42&(2)(4)\\
CenA-MM-Dw4       &             200.7583&       -41.7861&       3.91&(2)(4)\\
CenA-MM-Dw6       &             201.4875&       -41.0942&       3.61&(2)(4)\\
CenA-MM-Dw7       &             201.6167&       -43.5567&       3.38&(2)(4)\\
CenA-MM-Dw2       &             202.4875&       -41.8731&       3.60&(2)(4)\\
CenA-MM-Dw1       &             202.5583&       -41.8933&       3.63&(2)(4)\\
CenA-MM-Dw3       &     202.5875&       -42.1925&       4.61&(2)(4)\\
CenA-MM-Dw9       &             203.2542&       -42.5300&       3.81&(2)(4)\\
CenA-MM-Dw8       &             203.3917&       -41.6078&       3.47&(2)(4)\\\\
dw1315-45         &             198.9833&       -45.7506&       3.68*&(3)(4)\\
dw1318-44         &     199.7417&       -44.8947&       3.68*&(3)(4)\\
CenA-MM-Dw11      &         200.4167&   -43.0825&   3.68*&(3)(4)\\
dw1322-39         &     200.6333&       -39.9060&       3.68*&(3)(4)\\
dw1323-40c        &     200.9042&       -40.7214&       3.68*&(3)(4)\\
dw1323-40b        &             200.9792&       -40.8358&       3.68*&(3)(4)\\
CenA-MM-Dw12      &     201.0417&       -42.1397&       3.68*&(3)(4)\\
dw1323-40         &     201.2208&       -40.7614&       3.68*&(3)(4)\\
dw1326-37         &     201.5917&       -37.3856&       3.68*&(3)(4)\\
CenA-MM-Dw10      &     201.7042&       -43.0000&       3.68*&(3)(4)\\
dw1329-45         &         202.2917&   -45.1753&   3.68*&(3)(4)\\
CenA-MM-Dw13      &             202.4625&       -43.5194&       3.68*&(3)(4)\\
dw1330-38         &     202.6708&       -38.1675&       3.68*&(3)(4)\\
dw1331-40         &     202.8583&       -40.2631&       3.68*&(3)(4)\\
dw1331-37         &     202.8833&       -37.0581&       3.68*&(3)(4)\\
dw1336-44         &     204.1833&       -44.4472&       3.68*&(3)(4)\\
dw1337-44         &     204.3917&       -44.2186&       3.68*&(3)(4)\\
dw1337-41         &             204.4792&       -41.9031&       3.68*&(3)(4)\\
dw1341-43         &             205.4042&       -43.8547&       3.68*&(3)(4)\\
dw1342-43         &     205.6633&       -43.2553&       3.68*&(3)(4)\\
\\[-0mm]
\end{tabular}
\end{minipage} 
\tablefoot{Galaxies with unknown distances are denoted with a *. We adopted the distance of Cen\,A (3.68\,Mpc) for them. Galaxies that are members of Plane 1 or 2 are denoted with $^1$ or $^2$, respectively, according to T15.
 Galaxies denoted with (1) belong to the {T15 sample}, (2) are from the {LV sample},  (3)   belong to the {Candidate sample,}  and (4) are in the {Complete sample}.}
\end{table*}
\section{Transformation between coordinate systems}
To compare our results with those of T15 we need to transform the 3D positions of all sample galaxies from the equatorial system to the Cen\,A reference frame. This is carried out in three steps: (i) a transformation from equatorial (RA, DEC) to the Galactic ($l$, $b$) coordinates, (ii) a transformation to the supergalactic (SGL, SGB) coordinates, and (iii) a translation and rotation to the Cen\,A reference frame (CaX, CaY, CaZ). In the latter reference system, Cen\,A is located at the origin, and the two nearly parallel galaxy planes, represented by an averaged normal direction, lie in the XY projection. The two planes are then visible edge-on in the XZ projection (see our Sect.\,4 and Fig.\,2 in T15).

The transformation between polar coordinates and {Cartesian} coordinates is given by\\
$$x=d\cdot \text{cos}(\delta)\cdot \text{cos}(\alpha)$$
$$y=d\cdot \text{cos}(\delta)\cdot \text{sin}(\alpha)$$
$$z=d\cdot \text{sin}(\delta), $$
where $d$ is the heliocentric distance to the galaxy. To rotate from equatorial coordinates to Galactic coordinates, the Cartesian coordinates $v=(x,y,z)$ (equatorial system) have to be  multiplied from the left by the rotation matrix
$$\boldsymbol{R_{G}}=
 \begin{bmatrix}
-0.0549 &-0.8734 & -0.4839  \\
+0.4941 & -0.4448 & +0.7470  \\
-0.8677 & -0.1981 & +0.4560 \\
\end{bmatrix}.
$$ 
From galactic coordinates to supergalactic coordinates, the rotation is given by the matrix
$$\boldsymbol{R_{SG}}=
 \begin{bmatrix}
-0.7357 & +0.6773& +0.0000 \\
-0.0746 & -0.0810 & +0.9940\\
+0.6731 & +0.7313 & +0.1101  \\
\end{bmatrix}.
$$ 
Therefore the transformation from equatorial to supergalactic coordinates is:
$$v_{SG} = \boldsymbol{R_{SG}}\boldsymbol{R_{G}}v.$$ 

In T15 the authors transformed the supergalactic coordinates into the Cen\,A reference frame. To do this, they applied a translation to the coordinate system such that Cen\,A is at the origin (0,0,0)
$$
v_{SG,CenA} =  
v_{SG} +
\begin{pmatrix}
+3.41 \\
-1.26  \\
+0.33  \\
\end{pmatrix} \text{ [Mpc]}
$$  
and then rotate into the new coordinate system with
$$\boldsymbol{R_{Ca}}=
 \begin{bmatrix}
+0.994 & -0.043 & +0.102 \\
-0.001 & +0.919 & +0.393\\
-0.111 & -0.391 & +0.914  \\
\end{bmatrix}.
$$ 
The final transformation to the Cen\,A reference frame is
$$v_{Ca}=\boldsymbol{R_{Ca}}v_{SG,CenA.}$$
{We note that T15 uses 3.66\,Mpc as distance for Cen\,A, while their given translation from $v_{SG}$ to $v_{SG,CenA}$ only adds up to 3.65\,Mpc. In this work we use an updated distance value of 3.68\,Mpc taken from the LV catalog. These small differences place Cen\,A with a minor offset to the center of the reference frame.}
\section{Geometrical alignment of the satellites}

\begin{figure*}
\includegraphics[width=18cm]{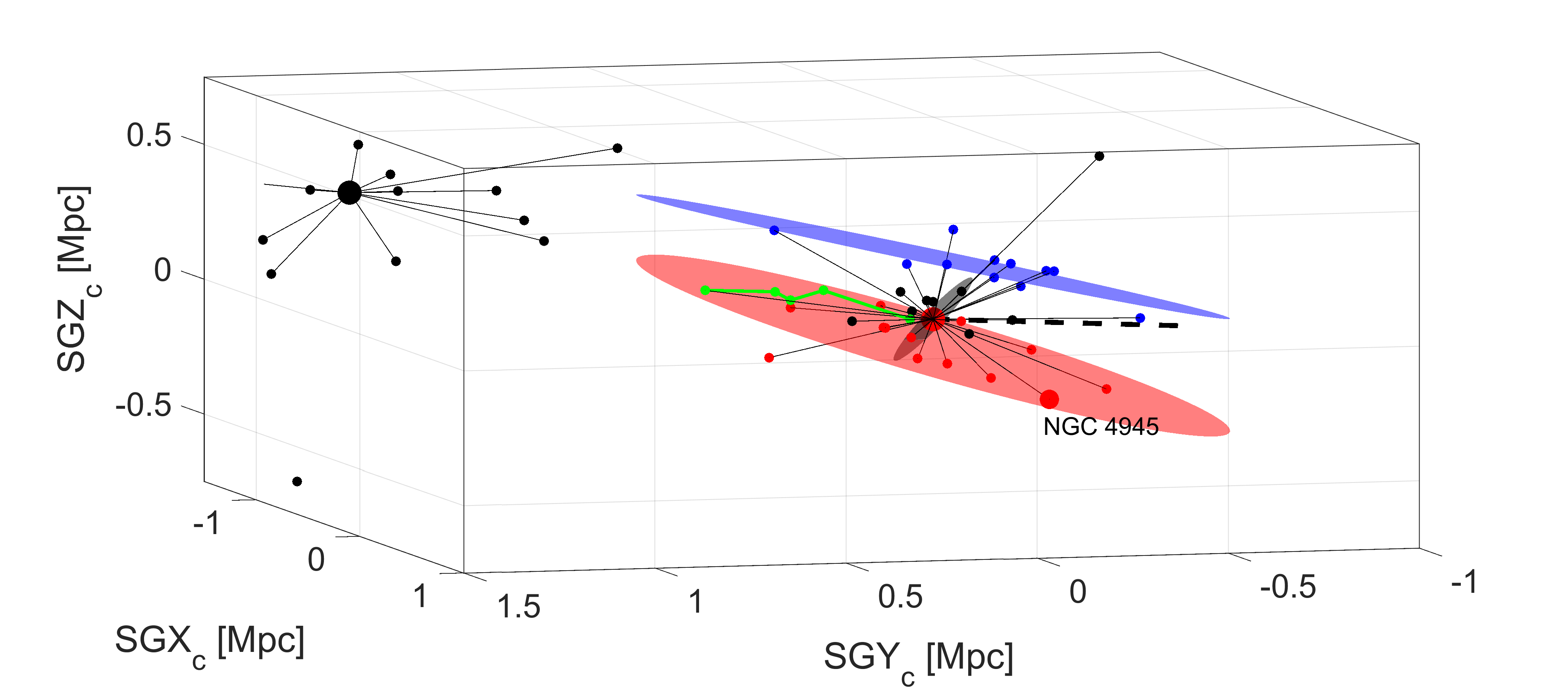}
\caption{Three-dimensional distribution of known Centaurus group members in supergalactic coordinates centered on Cen\,A. The large red dot is Cen\,A, the large black dot is M\,83, and the intermediate-size red dot is the late-type spiral NGC\,4945. The small red dots are plane 1 dwarf satellite members, the blue dots plane 2 members from T15, the black dots are additional (non-plane) members of the Cen\,A group. The best-fitting planes 1 and 2 are illustrated by the red and blue disks, respectively. The smaller gray disk that stands almost orthogonally to the two satellite planes represents the dust lane of Cen\,A. The dashed black line corresponds to the line of sight toward Cen\,A. The green line connects the five regions in the tail of the tidally disrupted dwarf CenA-MM-Dw3 where distances could be measured \citep[Dw3, Dw3\,S, Dw3\,SE, Dw3\,N, and Dw3\,NW, ][]{2016ApJ...823...19C}. 
}
\label{3dPlotNear}
\end{figure*}

We used two different algorithms to fit the two planes of satellites in three dimensions:
a singular value decomposition \citep[svd; ][]{1965SJNA....2..205G} and  the tensor of inertia \citep[ToI; e.g., ][]{2015ApJ...815...19P}. The 14 galaxies used for fitting plane 1 are labeled  $^1$ in Table \ref{table:2} and the 11 galaxies for plane 2 are labeled $^2$.  The normal vectors of the best-fitting planes given in supergalactic coordinates are $\mathbf{n_1} =(-0.1576,-0.4306,0.8886)$ and $\mathbf{n_2}=(0.0875,0.3225,-0.9425)$ for plane 1 and 2, respectively. Both algorithms return the same results. 
The angle between the two normal vectors is found to be 8.0$^{\circ}$, which is in good agreement with 7$^{\circ}$ published by T15. We cannot evaluate why there is a difference between our results and T15, as neither the plane-fitting method nor the exact sample of galaxies used for the fits were stated in T15. The measured angles between $\mathbf{n_1}$, $\mathbf{n_2}$, and 
the normal vector of the supergalactic plane are 16.9$^{\circ}$ and 24.6$^{\circ}$, respectively, meaning that the two planes are not far from being parallel to the supergalactic plane. Using the ToI algorithm we further calculated that plane 1 has a root-mean-square (rms) thickness of 69\,kpc and a major-axis rms length of 309\,kpc, while plane 2 has a thickness of 48\,kpc and a length of 306\,kpc. 

As the prominent dust lane of Cen\,A itself represents a reference plane, one can further ask how the satellite planes are spatially arranged with respect to Cen\,A. The orientation of the dust lane was studied by \cite{1995ApJ...449..592H} who give an inclination of 17$^{\circ}$ to the line of sight and a position angle for the disk angular momentum of 35$^{\circ}$ in the sky (counting from N through E). This latter angle 
also coincides with the position
angle of the photometric major
axis (Dufour et al.~1979). 
From these two angles we calculated the normal vector of the Cen\,A dust plane in supergalactic coordinates to be $\mathbf{n_{dust}} = (-0.0305, 0.8330, 0.5525)$. The resulting angular difference to the normal vectors of the satellite planes 1 and 2 amounts to 82.1$^{\circ}$ and 104.8$^{\circ}$, respectively. This means that the satellite planes are almost orthogonal to the dust plane of Cen\,A (see Fig.\,1), which is reminiscent of the local situation where the VPOS is essentially perpendicular to the plane of the Milky Way.   Furthermore, the angle between the connection line of Cen\,A and M\,83  and the normal of the dust plane is only 30$^{\circ}$, meaning that $\mathbf{n_{dust}}$ almost points toward M\,83.

\cite{2016ApJ...823...19C} discovered the tidally disrupted dwarf galaxy CenA-MM-Dw3 with tails spanning over 1$^{\circ}$.5 ($\sim 120$\,kpc). They measured distances at five separate high surface brightness locations along the elongation.  Defining the directional vector of CenA-MM-Dw3 as the vector between CenA-MM-Dw3 and  CenA-MM-Dw3-SE,
 the vector is $\mathbf{e_{Dw3}}=(0.8983,-0.4266, 0.1056)$ in supergalactic coordinates.  Astonishingly, we find that angles between this vector and the plane 1 and 2 normal vectors are  82$^{\circ}$  and 99$^{\circ}$, respectively, meaning that the tidally disrupted dwarf is almost parallel to the two planes. This is again reminiscent of the Local Group where the Magellan Stream aligns with the VPOS. In contrast to the Local Group the angle between $\mathbf{e_{Dw3}}$ and the normal of the dust plane is 109$^{\circ}$, meaning that the tidal dwarfs is almost parallel to the dust plane. The tail itself lies along the line of sight, 
 having the positive effect that the distance errors only move along this direction, hence the uncertainties do not change the angles between the tail and the planes. In Table\,\ref{table:vectors} we compile the calculated vectors in supergalactic coordinates.

\begin{table}[H]
\caption{Directions in supergalactic coordinates.}             
\label{table:vectors}      
\centering                          
\begin{tabular}{l l c}        
\hline\hline                 
 &  & (SGX, SGY, SGZ) \\    
\hline \\[-2mm]                 
normal plane 1 & $\mathbf{n_1}$ & $(-0.1576, -0.4306, +0.8886)$  \\ 
normal plane 2 & $\mathbf{n_2}$ & $(+0.0875,+0.3225,-0.9425)$  \\
normal Cen\,A dust plane & $\mathbf{n_{dust}}$ &$(-0.0305,+0.8330,+0.5525)$\\
elongation CenA-MM-Dw3 & $\mathbf{e_{Dw3}}$     & $(+0.8983,-0.4266, +0.1056)$  \\
\hline                                   
\end{tabular}
\end{table}

Fig.\,\ref{3dPlotNear} shows the 3D galaxy distribution of the Centaurus group in supergalactic coordinates with Cen\,A at the origin. Data are drawn from the online version of the LV catalog \citep{2013AJ....145..101K}. The primary double structure of the group is defined by the Cen\,A (big red dot) subgroup and the M\,83 (big black dot) subgroup, and the secondary double-plane structure around the Cen\,A subgroup. The 14 Plane 1 satellites are shown as red dots, the 11 Plane 2 satellites as blue dots according to the {T15 sample}. The intermediate-size red dot is NGC\;4945.  The best-fitting planes are also indicated (shown as red and blue disks; for details on the fitting procedure see above), the alignment of the tidal features of CenA-MM-Dw3 (green line with dots), and the dust plane of Cen\,A (in gray).

\section{Mapping the new dwarf candidates}

\begin{figure*}
\centering
\includegraphics[width=8cm]{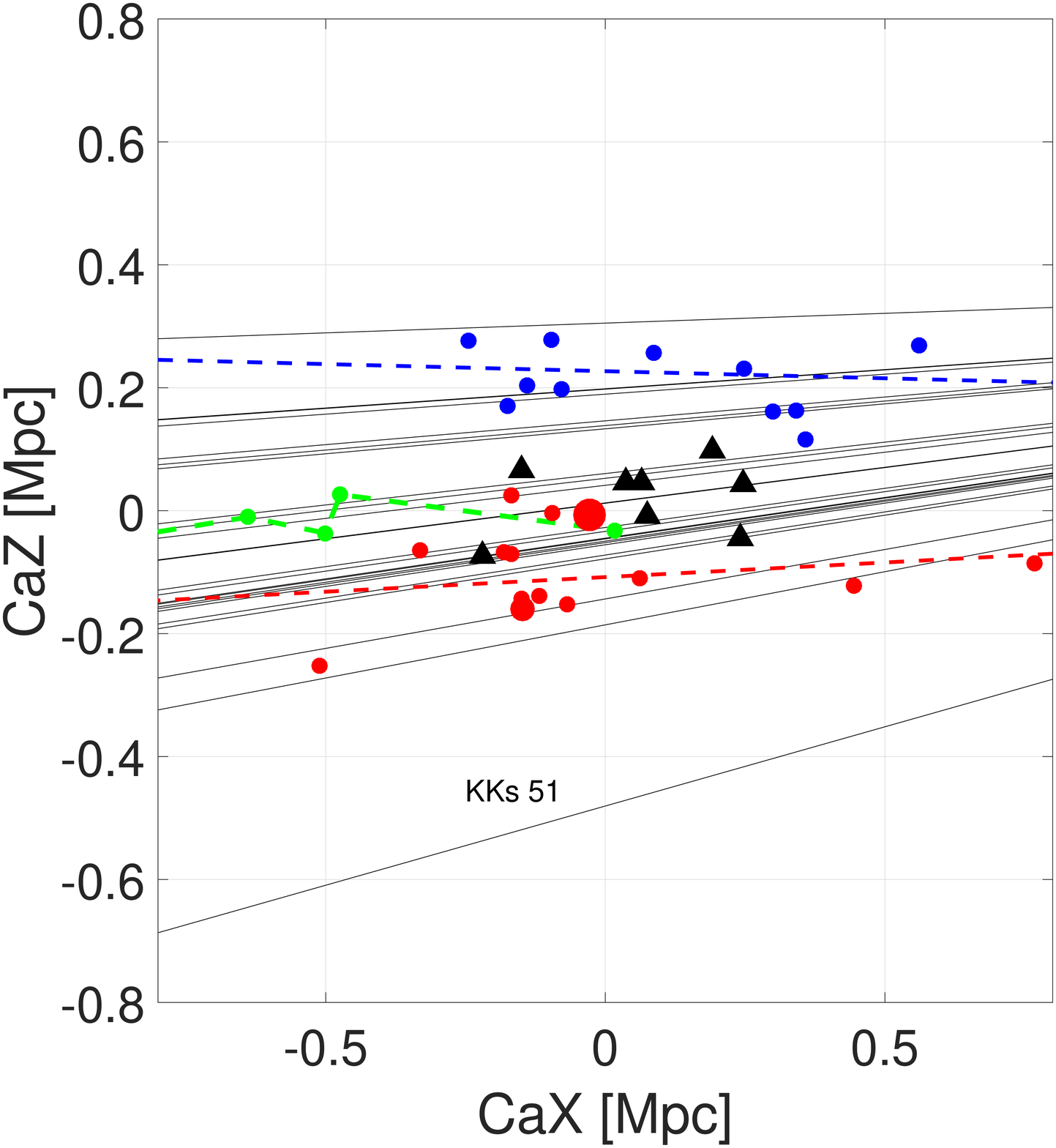}
\includegraphics[width=8cm]{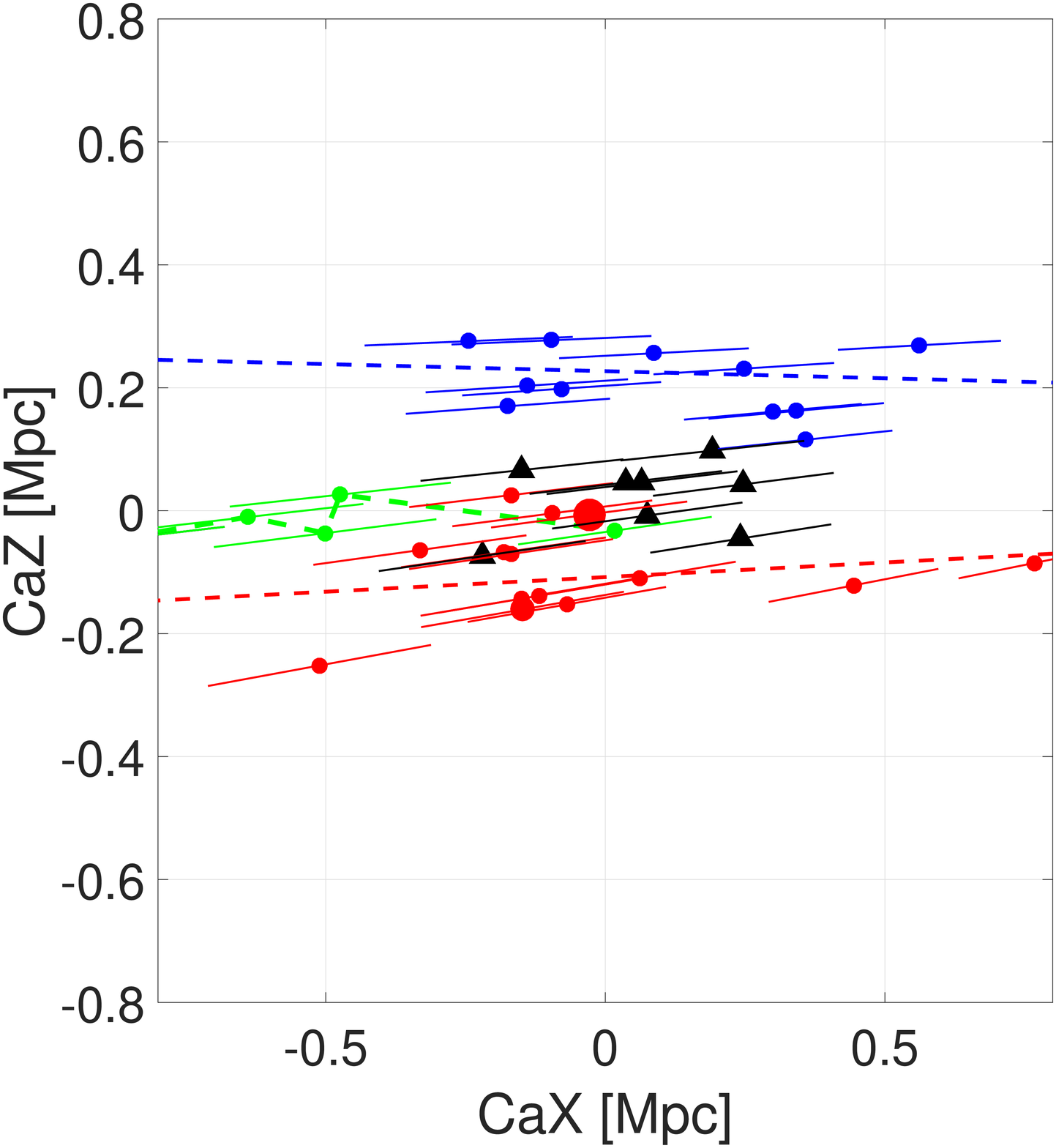}\\
\includegraphics[width=8cm]{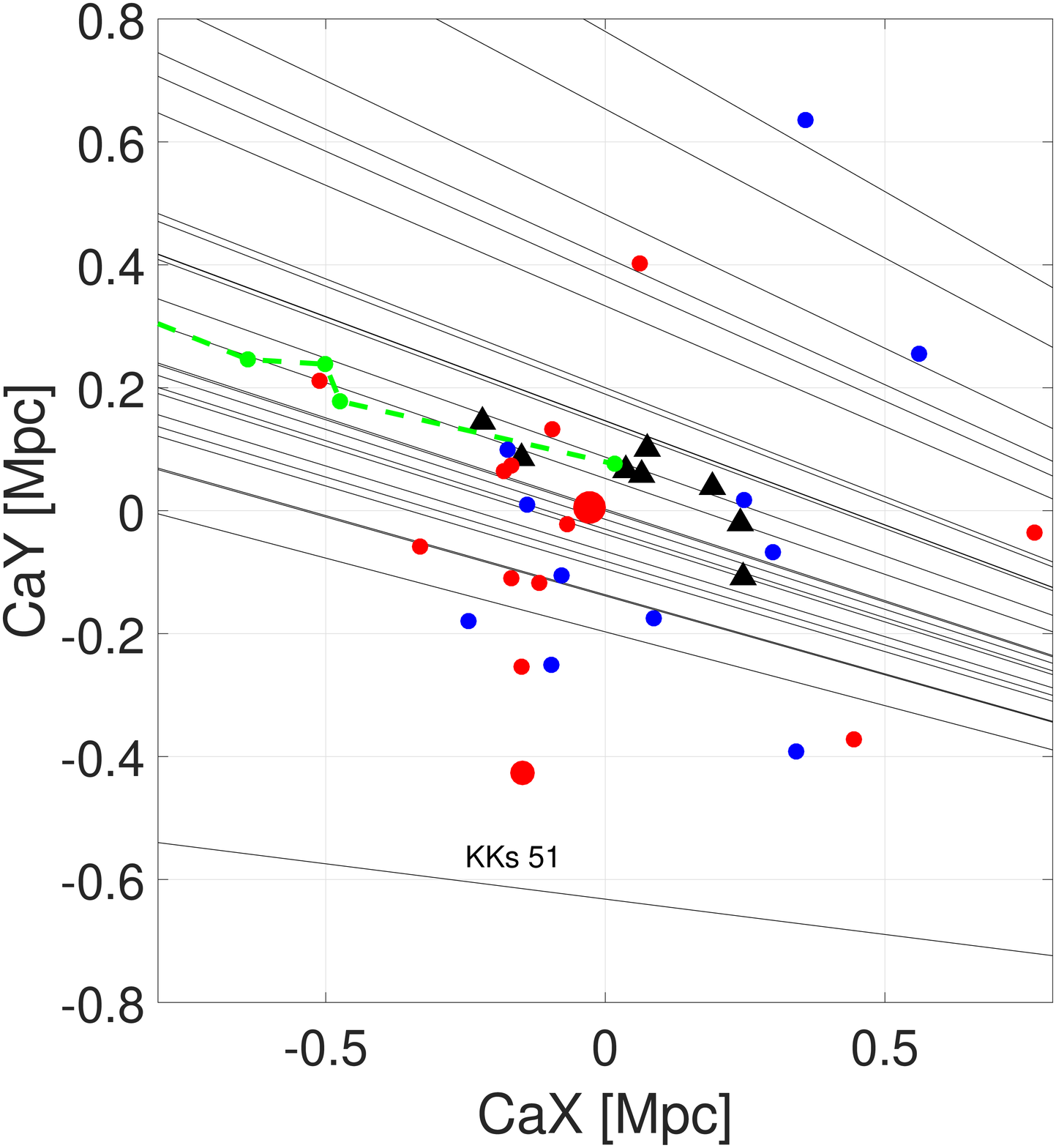}
\includegraphics[width=8cm]{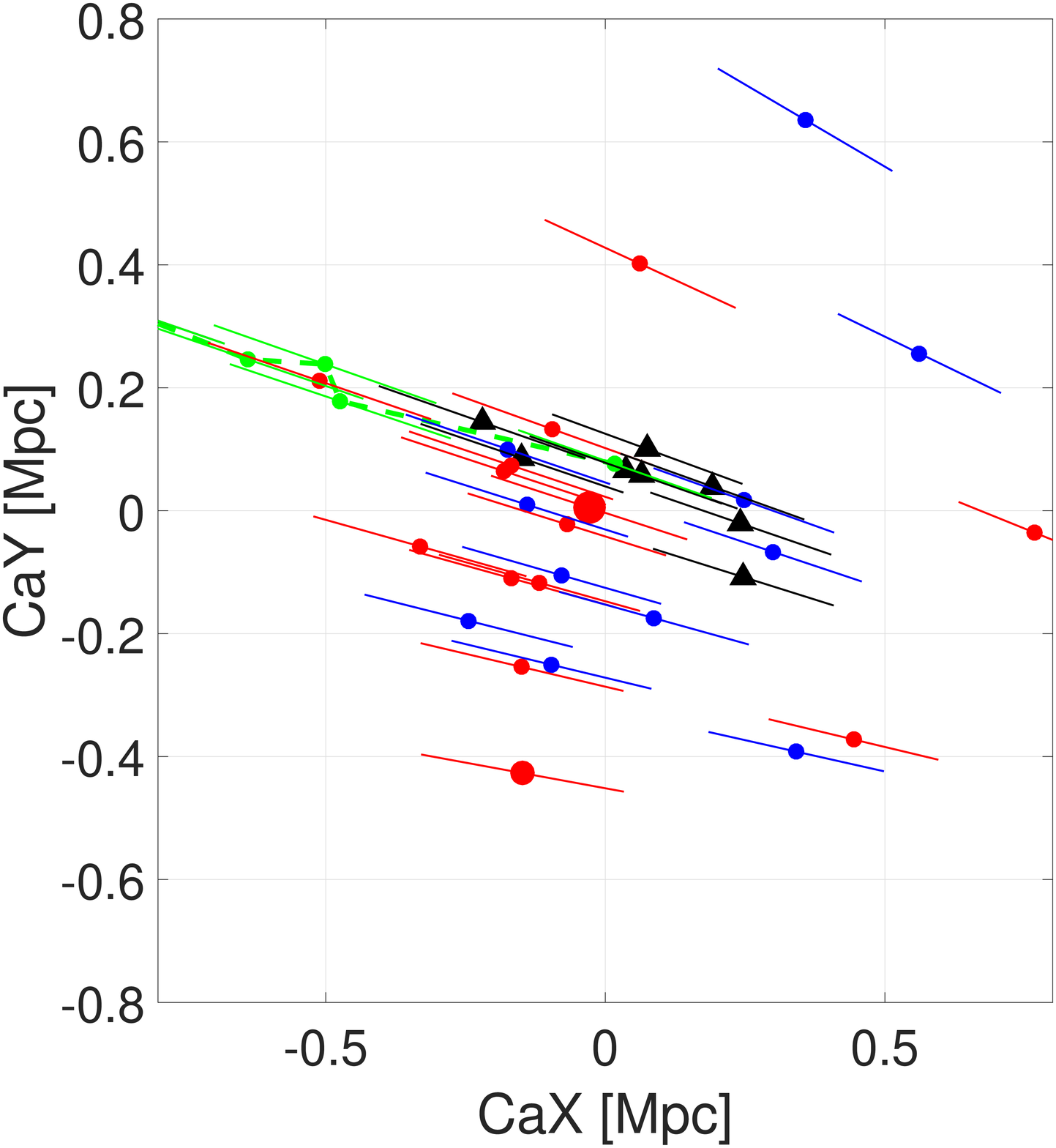}
\caption{Top left: edge-on view of the two galaxy planes. Red dots are satellites in plane 1, blue dots are satellites in plane 2, and black triangles are new dwarfs from the PISCeS survey. The green dotted line corresponds to the tidally disrupted dwarf CenA-MM-Dw3 where the dots themselves are regions in the tail where distances were measured (Dw3\,S, Dw3\,SE, Dw3\,N, and Dw3\,NW). CenA-MM-Dw3 itself falls outside the frame. The red and blue dashed lines are the best-fitting planes. The 25 thin black lines indicate the possible locations of new dwarf candidates without distance measurements. 
Top right: same as top left but without the possible group members, giving instead an indication of the distance uncertainties of the known members. The colored lines correspond to 5 percent distance errors, projected onto the CaX - CaZ plane. Bottom left: galaxy distribution in the CaX - CaY plane, where we see the two planes superimposed and face-on. Bottom right:  same as bottom left, only showing the galaxies with known distances. As in Fig.\,1, the large red dot is Cen\,A and the intermediate-size red dot is the giant spiral NGC\,4945.}
\label{tullyXZ}
\end{figure*}

We now switch to the Cen\,A reference frame (CaX, CaY, CaZ) that was introduced by T15. The top panels of Fig.\,\ref{tullyXZ} show the distribution of the Cen\,A group members in the CaX\,--\,CaZ projection; this plot is directly comparable to Fig.\,2 of T15. Here, the two galaxy planes are seen edge-on. The bottom panels show the CaX\,--\,CaY projection, presenting the planes in a face-on view. As has already been noted in T15, our line of sight lies almost in the direction of the two planes. The angle between the best-fitting planes and the line of sight at the distance of Cen\,A is only 14.6$^{\circ}$ and 12.8$^{\circ}$ (with a different sign), respectively, for plane 1 and 2. As a consequence of this geometry, any distance uncertainties move galaxies essentially along the two planes and thus have little bearing on the bimodal structure. On the other hand, for the same reason one might surmise that the planes are an artifact produced by the spread of distance errors. This is unlikely to be the case, however, as the planes are significantly more extended than the distance errors. T15 came to the same conclusion.

The special geometrical situation allows us to put the two planes of satellites to the test with the help of our new dwarf candidates around Cen\,A \citep{2016arXiv160504130M}. Given the celestial position (i.e.,~equatorial coordinates) of a dwarf candidate, its relative position to the planes is essentially given as well. There are three possibilities for a candidate: (1) it lies in one of the two planes, (2) it lies between the planes, or (3) it lies outside of the bimodal structure. Again, the key point is that this test can be conducted without knowing the distances of the candidates, deferring, at least in this preliminary manner, the {observationally challenging} task of distance measurements. 

In the top panels of Fig.\,\ref{tullyXZ} all known Cen\,A group members with distances are plotted in the CaX - CaZ projection. Red dots are the galaxies in plane 1, blue dots are galaxies in plane 2 according to T15, and black triangles are the newly found galaxies by \cite{2014ApJ...795L..35C,2016ApJ...823...19C}.
The best-fitting planes, i.e.,\,their lines of intersection with the orthogonal CaX - CaZ plane, are shown in the top left panel. In the top right panel the 5 percent distance uncertainties for the galaxies are added, showing the distance range along the line of sight. 
Likewise, the thin black lines in the top left panel indicate the lines of sight for the candidate Cen\,A group members,
nicely illustrating the near parallelism between the satellite planes and our line of sight. The bottom panels show the CaX - CaY projection, looking at the plane face-on. 

From the CaX - CaZ projection it becomes clear that the 25 dwarf galaxy candidates, provided they are Cen\,A group members, can be assigned almost unambiguously to one of the two planes because there is no double plane crossing along the line of sight at the distance of Cen\,A. In fact, of the 25 lines of sight (=\,possible positions of the candidates) none are crossing both planes within the distance range considered ($D_{Cen\,A}\pm$ 0.75\,Mpc); see Fig.\,\ref{tullyXZ}, top left. One candidate, KKs\,51, {clearly misses} both planes. For example, to be a member of plane 1, it would have to be 2\,Mpc
from the Milky Way (or 1.9\,Mpc from Cen\,A) where the line of sight and plane 1 intersect. 
In T15, this candidate is indicated as {other?}. We conclude that KKs\,51 is not a member of the planes and therefore should count as {other}. Looking solely at the CaX-CaZ projection, there remains the possibility that the lines of sight are only projected into the region of the planes, while in the 3D reality they could lie way off the planes. The CaX-CaY projection (face-on view) in the bottom panels shows that this is not the case. Only the line of KKs\,51 is far from the Cen\,A galaxy aggregation. 

Interestingly, the nine PiSCeS dwarfs (black triangles in Fig.\,2) seem to fill the gap between the two planes. This raises the question what will happen when the large sample of our new dwarf candidates comes into play. Will the bimodal structure be lost altogether, unmasking the double-plane structure of the Cen\,A subgroup as an effect of small-number statistics, or will the case for a double plane be strengthened?

\begin{figure}[ht]
\includegraphics[width=10cm]{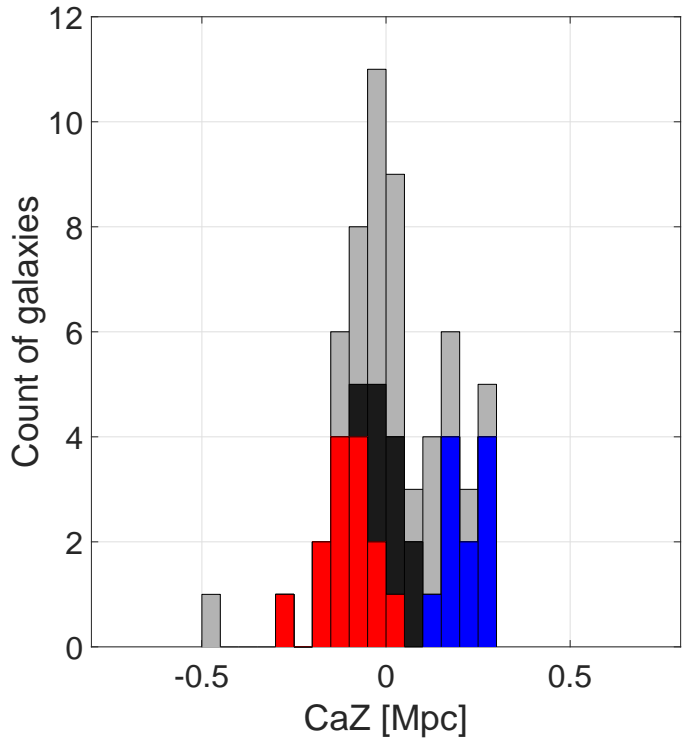}
\caption{Histogram of the Cen\,A satellite distribution along the CaZ axis. The red and blue bins correspond to the (14+11) plane 1 and 2 members, respectively, from the T15 sample. 
The gray bins are the 25 Cen\,A subgroup member candidates assuming a distance of 3.68\,Mpc (Cen\,A).}
\label{histogramAll}
\end{figure}

The Cen\,A subgroup members and candidates can now be sampled along the CaZ axis. In T15 this is carried out in their Fig\,2 by plotting a histogram of CaZ coordinates for members with their individually measured distances and for possible members (the dwarf candidates) assuming a distance of 3.68\,Mpc (the distance of Cen\,A). We binned our galaxy samples as in Fig\,\ref{histogramAll}. The red and blue bins correspond to the plane galaxies from the {T15 sample}, colored as before. 
The bimodal structure is clearly visible. If we add the nine satellites (in black) from the PISCeS survey \citep{2014ApJ...795L..35C,2016ApJ...823...19C} the gap starts to fill up. A limitation of the PiSCeS survey, however, is the relatively small area of 11\,deg$^{2}$ covered \citep[see Fig.\,2 in][]{2016arXiv160504130M}. Nevertheless, {despite} this possible bias, every dwarf that is found between the two planes reduces the significance of the bimodality. Finally, when including the more homogeneous and bias-free sample of 25 candidates (gray bins) that covers most of the vicinity of Cen\,A from the {Candidate sample} \citep{2016arXiv160504130M,2016ApJ...823...19C} the bimodality in the distribution increases slightly again and the population of plane 1 becomes dominant.

\begin{figure}
\includegraphics[width=10cm]{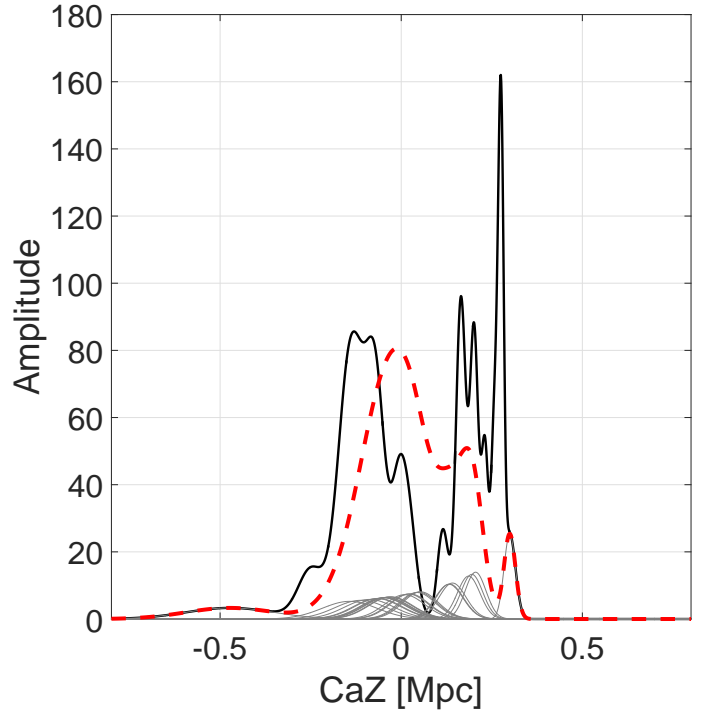}
\caption{Adaptive kernel for the Cen\,A subgroup members. Shown is the density distribution of satellite galaxies along the CaZ axis from a superposition of projected standard normal curves accounting for distance uncertainties (see text). The black line corresponds to the {T15 sample} (members and candidates), and the red dotted line to the {Candidate sample}, assuming a distance of 3.68 Mpc (Cen\,A) and an uncertainty due to the depth of the subgroup of $\pm$\,0.5\,Mpc. The standard normal curves of the candidates are shown in gray.
}
\label{adaptivKernelDiff}
\end{figure}

\begin{figure}
\centering
\includegraphics[width=10cm]{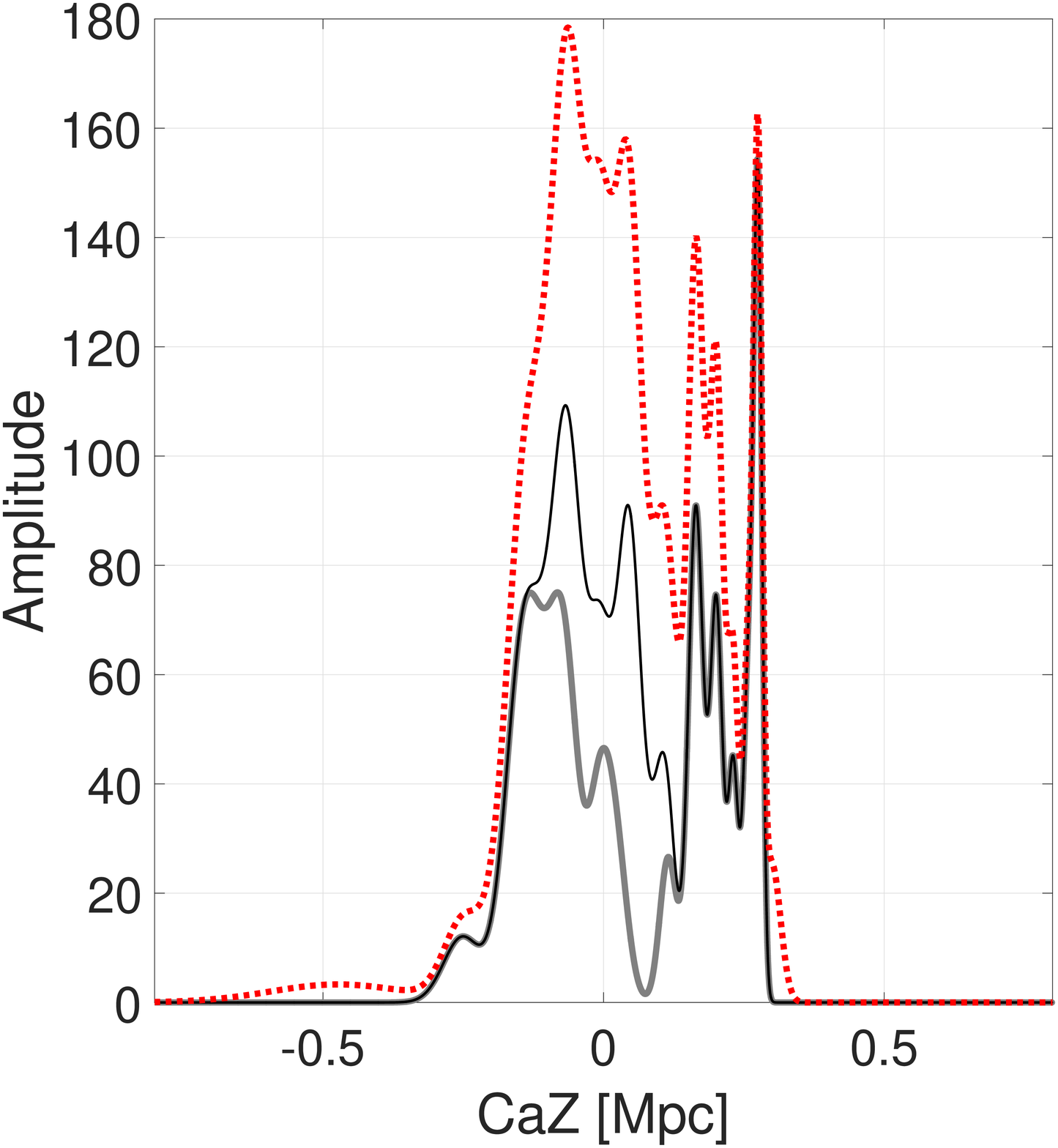}
\caption{Same as Fig.\,\ref{adaptivKernelDiff} but including the PISCeS dwarfs and shown as {density} distribution, analoguous to the histogram of 
Fig.\,\ref{histogramAll}. The gray line corresponds to the members of the two planes from the {T15 sample}. The black line corresponds to the {LV sample}, comprising the {T15 sample} plus the nine new dwarfs from the PISCeS survey, thus representing all known galaxies with distances $<1$\,Mpc from Cen\,A. The red dotted line is the grand total, i.e.,\,the superposition of the black line and all the possible members of the Cen\,A subgroup from the {Candidate sample} \citep{2016arXiv160504130M,2016ApJ...823...19C}, assuming a distance of 3.68 Mpc (Cen\,A) and an error of $\pm$\,0.5\,Mpc. The galaxy Cen\,A itself is located at CaZ=0.}
\label{adaptivKernelAll}
\end{figure}

However, this representation of the CaZ distribution as a histogram is somewhat problematic. Not only does a histogram depend on the chosen starting point and bin width, we also adopted the same fixed distance for all candidates. Although there is only a weak dependence of CaZ on the distance, owing to the fact that the line of sight is almost perpendicular to the CaZ axis, the mean CaZ range covered by a group member candidate is around 0.11\,Mpc in the distance interval $2.93<D<4.43$\,Mpc, which can shift a candidate back and forth by up to two bins of 0.05\,Mpc width. A better representation of the data can be achieved using the adaptive kernel density estimation. In this technique, the galaxies (members and candidates) are represented by standard normal curves with   
$\mu$ as the measured or assumed distance and $\sigma$ as the distance uncertainty amounting to 5\,percent for members, accounting for measurement errors, and 0.5\,Mpc for candidates, accounting for the depth of the subgroup. The standard normal curves are then projected onto the CaZ axis, resulting in standard normal curves with different $\sigma$ values, which depend on the projection angle, i.e.,\,the angle between the line of sight and CaZ. This angle varies systematically over CaZ (see Fig.\,\ref{tullyXZ}, top left), from close to orthogonal at positive CaZ values (CaZ$\sim 0.2$\,Mpc), giving very narrow standard curves, to ever smaller acute angles toward negative CaZ values (CaZ$\sim -0.3$\,Mpc), giving broader standard curves.  
The combined density distribution of the galaxies along the CaZ axis, by co-adding the projected normal curves, is given in Figs.\,\ref{adaptivKernelDiff} and \ref{adaptivKernelAll}. 

In Fig.\,\ref{adaptivKernelDiff} the {T15 sample} (black solid line) and all candidate galaxies without distances from the {Candidate sample} (red dashed line) are plotted differentially. For the galaxies without distance measurements a distance of 3.68\,Mpc and a distance uncertainty of $\pm 0.5$\,Mpc is assumed. These standard normal curves are plotted in gray. The bimodality is visible, albeit the peak around plane 1 is higher than the peak around plane 2.   

Fig.\,\ref{adaptivKernelAll} shows the same data as the histogram in Fig.\,\ref{histogramAll}. This plot contains all the information up to date. The gray line corresponds to the plane satellites from the {T15 sample}, the black line to the {LV sample,} and the red dashed line to the {Complete sample}. In all three samples the gap is visible. It is now important to test the significance of the planar structures. 
\section{Statistical analysis}
Is the seemingly bimodal structure of the Cen\,A subgroup statistically significant? To answer this question we performed two different statistical tests,  the Anderson-Darling test \citep{1986gft..book.....D} and the Hartigan dip test \citep{10.2307/2241144}.
The null hypothesis is that the observed galaxy distribution can be described by a unimodal normal distribution.   
Basically, we ask whether or not the galaxy distribution in the edge-on view, sampled along the CaZ axis, can be explained by a single normal distribution. If the null hypothesis is rejected, there must be evidence of multimodality in the Cen\,A subgroup, or the assumption of an underlying normal distribution is incorrect. We test the four samples listed in Table\,1: (1) the {T15 sample}, (2) the {LV sample}, (3) the {Candidate sample}, and (4) the {Complete sample}. 

The Anderson-Darling test is an improved version of the Kolmogorov-Smirnov test. Both of these tests compare the standard normal cumulative distribution function (CDF) with the empirical cumulative distribution function (ECDF) created from the data. If the difference between the ECDF and CDF becomes larger than a critical value, the hypothesis of normality is rejected with some significance level. In contrast to Kolmogorov-Smirnov, the Anderson-Darling test weights the tails of a distribution higher than the center. We use the implementation of the Anderson-Darling test provided by MATLAB.

The results for the Anderson-Darling test are presented in Table\,4: samples (2), (3), and (4) pass the test for normality at the significance level of  5\% with $p_{AD}$-values of 0.18, 0.07, and 0.15, respectively. With a $p_{AD}$-value of 0.07 the candidates-only sample (3) has still a marginal probability of not being unimodally distributed.
Sample (1) is inconsistent with being drawn from a normal distribution with a $p_{AD}$-value of 0.02, an expected result given the finding of T15. 
Taking all the recently added data into account, however, the distribution of galaxies along the CaZ axis becomes consistent with being normally distributed.

There is a possible caveat. The footprint of the PISCeS survey does not cover the whole area on the sky taken by the two planes, see Fig.\,2 of \cite{2016arXiv160504130M}, which could lead to a selection bias. Disregarding the findings of the PISCeS survey: Dw1, Dw2, Dw4, Dw8, and Dw9 would have been detected in the data of our Centaurus group survey, but not the extremely faint and diffuse objects Dw5, Dw6, and Dw7. Furthermore, the photometric properties of the tidally disrupted dwarf Dw3 does not conform with our search criteria and would have remained undetected as well. We also assume that Dw10, Dw11, Dw12, and Dw13 would not have been detected (no photometry was performed in \cite{2016ApJ...823...19C} for these candidates). In total, only five Dw galaxies would then feature in the {Candidate sample} and {Complete sample} with an assumed distance of 3.68\,Mpc. Do the significance tests return different results with these conservative assumptions? We repeated the Anderson-Darling test with the following results: the null hypothesis is rejected for samples (1) and (2), while samples (3) and (4) are consistent with a normal distribution at the 5\% confidence level. Sample (2) is rejected because it now contains only the galaxies with distances from the {T15 sample}. 
We therefore conclude that the outcome of the Anderson-Darling is little affected by a selection bias from including the PISCeS dwarfs in the analysis.

The Hartigan dip test provides another method to test for bimodality. It measures the maximum difference between the empirical distribution function and the unimodal distribution function that minimizes that maximum difference. We use an implementation originated by \cite{10.2307/2347485}, which was translated into MATLAB by Ferenc Mechler. 

The results for the Hartigan dip test are as follows: samples (2), (3), and (4) pass the test at the significance level of 5\% with p-values of 0.61, 0.72, and 0.84, respectively. Only sample (1) fails the test with a p-value of 0.00. This is the same result as before. Again we checked for a change of the results by the exclusion of the five candidates from the {T15 sample}. With a $p$-value of 0.08 the result for this plane-members-only subsample changes indeed. It now has a marginal probability of not being unimodally distributed, and when testing at a 10\% significance level it is clearly rejected from unimodality. Overall the conclusion is hence the same: while the galaxy data used in the original analysis by Tully and collaborators support the picture of a multimodal distribution (two planes), adding the new galaxies, and thereby doubling the sample, presents a picture that is consistent with a unimodal normal distribution. 

Until now no distance uncertainties were taken into account when testing for the significance of the planes. To test whether these results are indeed representative of the data and their uncertainties, we performed Monte Carlo simulations where the distance of a galaxy is randomly taken from a normal distribution with $\mu$ given by the galaxy position and $\sigma$ by the distance error of 5\,\%. Remember that we sample along the CaZ axis, which means that we only count the CaZ component of the distance uncertainty. For galaxies without distance measurements a conservative distance uncertainty of $\pm$0.5\,Mpc, accounting for the depth of the Cen\,A subgroup, is taken. We calculated 10000 realizations of all samples and applied both the Anderson-Darling and the Hartigan dip test on them. The resulting probabilities that the null hypothesis (=\,normal distributed) is rejected for the Anderson Darling and the Hartigan dip test are as follows: 
(1) is rejected in 66 and 62 percent of the draws, (2) in 2 and 0 percent, (3) in 21 and 0 percent, and (4) in 5 and 0 percent. 

In this analysis all candidates are considered to be satellites of Cen\,A. But what if some of these candidates are not members of the Cen\,A subgroup but lie in the background? To check this possibility, we broadened the distance uncertainty to $\pm$1.5\,Mpc and repeated the MC runs. Remember that we exclude all galaxies with radial distances to Cen\,A larger than 1\,Mpc. 
When we rerun the simulations with this setup, (1) gets rejected in 65 and 56 percent of the cases, (3) in 18 and 2 percent, and (4) in 5 and 0 percent. Hence this wider spread essentially lowers the rejection rate.

The results of these tests are summarized in Table\,\ref{table:3}. Column 1 lists the  test sample name. Columns 2 and 3 gives the accepted hypothesis and $p$-value from the Anderson-Darling test.  Column 4 lists the\ probability for a rejection of h$_0$ estimated with the Anderson-Darling test from Monte Carlo simulations.  The results from the test with a distance uncertainty of $\pm$\,1.5\,Mpc are indicated in brackets. Columns 5 and 6 gives the accepted hypothesis and $p$-value from the Hartigan dip test. Column 7 lists the probability for a rejection of h$_0$ estimated with the Hartigan dip test from Monte Carlo simulations. In brackets The results from the test when adopting a distance uncertainty of 1.5\,Mpc are indicated in brackets.

\begin{table}[H]
\caption{Anderson-Darling and Hartigan dip test results.}             
\label{table:3}      
\centering                          
\begin{tabular}{l c c l c c l}        
\hline\hline                 
Sample & h$_{ad}$ & $p_{AD}$ & P$_{AD,MC}$ & h$_{dip}$ & $p$ & P$_{dip,MC}$\\    
\hline       \\[-2mm]                 
   (1)                  & h$_1$ & 0.02 & 0.66 (0.65) &h$_1$ & 0.00 & 0.62 (0.56) \\      
   (2)                  & h$_0$ & 0.18 & 0.02 &h$_0$ & 0.61 & 0.00  \\
   (3)                  & h$_0$ & 0.07 & 0.21\,(0.18) &h$_0$ & 0.72 & 0.01\,(0.02)\\
   (4)                  & h$_0$ & 0.15 & 0.05\,(0.06) &h$_0$ & 0.84 & 0.00\,(0.00)\\
\hline                                   
\end{tabular}
\tablefoot{The value h$_0$ means that the null hypotheses of normal distribution passes the test, and h$_1$ means it is rejected. It is rejected when $p$\,<\,0.05. } 
\end{table}

From all our results, we conclude that with the addition of the new data the significance against a unimodal distribution of the satellites around Cen\,A rises, i.e.,\,the significance {for} bimodality is weakened, even though the case for a double-plane structure, as suggested by T15, is not completely ruled out. 

We performed one last test. What would happen if the new candidates were lying exactly  in (one of) the two planes? To find out,
we calculated the intersection point between the line of sight and the planes, and thus a new (hypothetical) distance $d_{inter}$ for each candidate from the {Candidate sample} and the {Complete sample}. To account for the thickness of the planes, we also calculated the intersection points at $\pm$ the r.m.s. thickness of the planes, giving a minimal ($d_{min}$) and maximal ($d_{max}$) value for the distance. We only take galaxies into account where the radial distance between the galaxy and Cen\,A is less than 1\,Mpc. In Table\,\ref{table:4} we present the estimated distances for all galaxies, where at least one of the three intersection points is closer than 1\,Mpc to Cen\,A. Distances indicated with an asterisk are outside of this 1\,Mpc radius. The following eight candidates miss the two planes in the 1\,Mpc vicinity of Cen\,A and are therefore removed from the modified {Candidate} and {Complete samples}: dw1329-45, CenA-MM-Dw13, dw1331-40, dw1337-41, KK198, KKs54, KKs59, and KKs51. 
\begin{table}[H]
\caption{Predicted distances for the candidates.}             
\label{table:4}      
\centering                          
\begin{tabular}{l l l l}        
\hline\hline                 
Name & d$_{\text{inter}}$  [Mpc] & d$_{\text{min}}$ [Mpc] & d$_{\text{max}}$ [Mpc]\\    
\hline       \\[-2mm]                 
Plane 1 & & & \\
dw1315-45               & 3.84  & 3.53  & 4.15  \\
dw1318-44      & 4.07   & 3.75  & 4.40  \\ 
CenA-MM-Dw11    & 4.15  & 3.82  & 4.45  \\
dw1322-39       & 3.83  & 3.52  & 4.13  \\
dw1323-40c      & 4.07  & 3.75  & 4.40  \\
dw1323-40b      & 4.13  & 3.80  & 4.46  \\
CenA-MM-Dw12    & 4.37  & 4.02  & 4.72* \\
dw1323-40       & 4.26  & 3.92  & 4.60  \\
dw1326-37       & 4.01  & 3.70  & 4.33  \\
CenA-MM-Dw10    & 4.99* & 4.59  & 5.39* \\ 
dw1330-38       & 4.86* & 4.47  & 5.25* \\
dw1331-37       & 4.85* & 4.46  & 5.23* \\
\\
Plane 2 & & & \\
dw1336-44               & 2.26* & 1.78* & 2.74  \\
dw1337-44       & 2.36* & 1.86* & 2.87  \\
dw1341-43       & 3.29  & 2.59* & 4.00  \\
dw1342-43       & 3.51  & 2.76  & 4.26  \\
KKs58                   & 3.02  & 2.38* & 3.67  \\
\hline                                   
\end{tabular}
\end{table}
Applying the Anderson-Darling and Hartigan dip tests to the modified {Candidate sample} and {Complete sample} leads to the following changes: (3) and (4) are now rejected by the Anderson-Darling test both with a $p_{AD}$-value of 0.00. The same is true when testing with $d_{min}$ and $d_{max}$. On the other hand, the Hartigan dip test still accepts the null hypotheses for samples (3) and (4), as before, but the $p$-values are lowered to 0.08 and 0.66, respectively. Again, we get the same result when testing with $d_{min}$ and $d_{max}$ instead of $d_{inter}$. With a $p$-value of 0.08, the modified {Candidate sample} is now marginally significant that is not unimodal, whereas the modified {Complete sample} gives a strong hint for unimodality, in contrast to the Anderson-Darling test. 

Hypothetically, putting the new candidates onto the best-fitting planes must of course strengthen the case for the planes. But as we cannot exclude the possibility that the dwarfs are indeed lying in the planes without accurate distance measurements, we also cannot conclusively rule out the reality of the planes.

\section{Discussion and conclusions}
The discovery of a large number of new dwarf galaxies in the nearby Centaurus group \citep{2014ApJ...795L..35C,2016ApJ...823...19C,2015A&A...583A..79M,2016arXiv160504130M} opened the opportunity to conduct a significance test of the two planes of satellites reported by \cite{2015ApJ...802L..25T}. While this normally requires follow-up observations to measure galaxy distances, in the case of the Cen\,A subgroup, owing to the special geometric situation, one can take advantage of the fact that the line of sight from our vantage point runs along the postulated planes of satellites. Therefore, galaxies even without distance measurements can be used for the test, as distance uncertainties move galaxies along or parallel to the planes. In other words, distance uncertainties produce only little crosstalk between the planes and the space around them. This allows us to include all Cen\,A member candidates in the analysis, which doubles the sample size from 30 to 59.

Sampling galaxy positions along an edge-on projection of the planes, we studied the distribution of Cen\,A subgroup members by two different techniques: a histogram and a more sophisticated adaptive kernel density estimation. A gap, or dip in the distribution marking the two planes, is visible in both representations of the data. However, it is also very evident that with the inclusion of the new galaxy data the gap between the two planes starts to be filled, raising the conjecture that the two plane scenario around Cen\,A might be an artifact of low number statistics. To put this under statistical scrutiny, we performed an Anderson-Darling test and a Hartigan dip test to see whether the distribution is in agreement with a unimodal normal distribution. We find that both tests fail with the sample used by \cite{2015ApJ...802L..25T}, rejecting the unimodal normal hypothesis for that orginal sample. This result is consistent with their finding that the satellites can be split into two planes. However, with the addition of the new dwarf members and candidates of the Cen\,A subgroup, the deviation from a normal distribution loses statistical significance in the sense of the two tests applied. Hence it is now conceivable that the satellites follow a normal distribution and the gap between the two planes is indeed an artefact from small number statistics. We performed Monte Carlos simulations to further strengthen these results by taking distance uncertainties into account and find that the results change only marginally. 

Given that distance measurements for the candidate galaxies are unavailable at the moment, it is theoretically possible to allocate 17 out of 25 galaxies to one of the two planes each by moving them along the line of sight.
That means that technically the existence of the two planes cannot be completely ruled out at this point. Only distance measurements will tell whether the candidates lie on or near the planes. All we can say is that the case for two planes around Cen\,A is weakened by including the currently available data for 29 new dwarfs and dwarf candidates.   

At first glance, another  secondary result of our testing is that in parallel with the weakening of the significance for bimodality, the galaxy population in plane 1 has now become dominant. This is not surprising: Cen\,A lies closer to plane 1 in projection (CaZ=0 in Fig.\,5), such that any newly discovered satellite galaxy in a distribution that is radially concentrated on Cen\,A will necessarily result in an additional member of plane 1. However, this amassing of Cen\,A satellites in the proposed plane 1 is important, as the planarity of Plane 1 has not been destroyed by adding the candidates. The alignment of the tidally disrupted CenA-MM-Dw3 with plane 1 (see Figs.\,1 and 2) is intriguing too. The Cen\,A plane 1 is certainly a good candidate analog of the local thin planes detected around the Milky Way (VPOS) and the Andromeda galaxy (GPOA). Future distance measurements for the many new Cen\,A subgroup member candidates will be able to tell how far the analogy will take us.   

\begin{acknowledgements}
OM, HJ, and BB are grateful to the Swiss National Science Foundation for financial support. HJ acknowledges the support of the Australian Research Council through Discovery projects DP120100475 and DP150100862. MSPs contribution to this publication was made possible through the support of a grant from the John Templeton Foundation. The opinions expressed in this publication are those of the authors and do not necessarily reflect the views of the John Templeton Foundation. {The authors like to thank the anonymous
referee for helpful comments that improved the paper.}
\end{acknowledgements}

\bibliographystyle{aa}
\bibliography{bibliographie}
\end{document}